\documentclass[conference]{IEEEtran}
\IEEEoverridecommandlockouts
\usepackage{cite}
\usepackage{amsmath,amssymb,amsfonts}
\usepackage{algorithmic}
\usepackage{graphicx}
\usepackage{textcomp}
\usepackage{xcolor}
\usepackage{amsmath,amsfonts}
\usepackage{algorithmic}
\usepackage{graphicx}
\usepackage{textcomp}
\usepackage{xcolor}
\usepackage{microtype}
\usepackage{mathptmx}
\usepackage{graphicx}
\usepackage{subcaption}
\usepackage{color}
\usepackage{relsize}
\usepackage{array}
\usepackage{makecell}
\usepackage{multirow}
\usepackage{booktabs}
\usepackage{pifont}
\usepackage{listings}
\usepackage{url}
\usepackage{paralist}
\def\BibTeX{{\rm B\kern-.05em{\sc i\kern-.025em b}\kern-.08em
    T\kern-.1667em\lower.7ex\hbox{E}\kern-.125emX}}

\begin{document}

\title{Auditing Gender Analyzers on Text Data}

\author{\IEEEauthorblockN{Siddharth D Jaiswal}
\IEEEauthorblockA{\textit{Dept. of CSE} \\
\textit{IIT Kharagpur, India}\\
siddsjaiswal@kgpian.iitkgp.ac.in}
\and
\IEEEauthorblockN{Ankit Verma}
\IEEEauthorblockA{\textit{Dept. of CSE} \\
\textit{IIT Kharagpur, India}\\
ankitverma@kgpian.iitkgp.ac.in}
\and
\IEEEauthorblockN{Animesh Mukherjee}
\IEEEauthorblockA{\textit{Dept. of CSE} \\
\textit{IIT Kharagpur, India}\\
animeshm@cse.iitkgp.ac.in}
}

\maketitle

\begin{abstract}
AI models have become extremely popular and accessible to the general public. However, they are continuously under the scanner due to their demonstrable biases toward various sections of the society like people of color and non-binary people.
In this study, we audit three existing gender analyzers -- \textsc{uClassify}, \textsc{Readable} and \textsc{HackerFactor}, for biases against non-binary individuals. These tools are designed to predict only the \textcolor{black}{cisgender} binary labels, which leads to discrimination against non-binary members of the society. We curate two datasets -- Reddit comments (660k) and, Tumblr posts (2.05M) and our experimental evaluation shows that the tools are highly inaccurate with the overall accuracy being $\approx 50\%$ on all platforms. Predictions for non-binary comments on all platforms are mostly \textit{female}, thus propagating the societal bias that non-binary individuals are effeminate. To address this, we fine-tune a BERT multi-label classifier on the two datasets in multiple combinations, observe an overall performance of $\approx 77\%$ on the most realistically deployable setting and a surprisingly higher performance of $90\%$ for the non-binary class. We also audit \textsc{ChatGPT} using zero-shot prompts on a small dataset (due to high pricing) and observe an average accuracy of 58\% for Reddit and Tumblr combined (with overall better results for Reddit).

Thus, we show that existing systems, including highly advanced ones like ChatGPT are biased, and need better audits and moderation and, that such societal biases can be addressed and alleviated through simple off-the-shelf models like BERT trained on more gender inclusive datasets.\footnote{\textcolor{red}{This work has been accepted at IEEE/ACM ASONAM 2023. Please cite the version appearing in the ASONAM proceedings.}}
\end{abstract}


\section{Introduction}
\label{sec:intro}
Gender, a characteristic defined by the socio-cultural structures that a person lives in, plays a huge role in how an individual is perceived in society and the kind of societal facilities that are made available to them. While there is a growing acceptance of the non-binary gender~\cite{richards2016genderqueer} and acknowledgement that it is different from the sex assigned at birth, most 
societies still treat gender as a binary attribute with only two identities- `male' and `female'~\cite{ukgovt2004}\footnote{In this work, we refer to cisgender men and women as male and female, respectively.}. 
This prejudiced practice that has been continuing for thousands of years has discriminated against various non-binary genders, and denied them access to basic public facilities like public toilets~\cite{weinhardt2017transgender,bagagli2021}, employment opportunities~\cite{bates2021employment} and civil rights, leading to legal discrimination~\cite{ungender2021}. 

\noindent \textbf{Impact of discrimination}: Societal discrimination can be dealt with through legal recourse, but overcoming these issues is arduous when there are non-human elements in the loop, like AI-based software, which are often \textit{unexplainable} black boxes.
Significant research has been done on discrimination against cisgender individuals in tasks like face recognition~\cite{buolamwini2018gender,jaiswal2022twoface}, automated hiring~\cite{suhr2021hiring}, and image search~\cite{feng2022has}, but there has been a very limited research~\cite{keyes2018misgendering,Scheuerman2019gender,jaiswal2022marching} investigating discrimination against non-binary individuals. A possible reason is the non-availability of `non-binary' as a gender label on many AI services like face recognition softwares~\cite{aws_rekognition,facepp,clarifai,microsoft_face} and gender analyzers from text~\cite{uclassify,readable,hackerfactor}. It becomes difficult to study the outputs for non-binary individuals under real deployment scenarios and erodes their existence from societal conscience. Questions can be raised on the morality of having a non-binary gender label but as these services are used at scale and deployed around the world, their impact on non-binary individuals is high. For example, a non-binary individual may be incorrectly classified as a male or female by a face recognition service and thus may be denied access to public toilets.

\begin{table}[t]
    \tiny
    \centering
	\begin{center}
		\begin{tabular}{|l |c | c| c | c | c|} 
            \hline
			\textbf{Platform} & \textbf{Comments} & \textbf{\textsc{UC}} & \textbf{\textsc{RD}} & \textbf{\textsc{HF}} & \textbf{\textsc{BT}}\\
			\hline
            Reddit & \makecell[l]{You're allowed to use he/him\\ AND they/them ... I use she/her \\ and he/him but not they/them ...\\  You're allowed to present\\ however you  want but still identify\\ as genderqueer, genderfluid,\\ non-binary, etc. ...} & F & F & WF & NB\\
			\hline
            Tumblr & \makecell[l]{“There is no queer community” is\\ not and will never be a true statement. \\ I am queer and I am part \\of a queer community. We are real,\\ we exist in real life, we are here. ... \\Because that is what you are.} & F & F & WF & NB\\
			\hline
		\end{tabular}
	\end{center}
 \vspace{-4mm}
	\caption{\footnotesize Some example comments by non-binary authors from Reddit and Tumblr dataset and the predicted labels by the gender analyzers. M: male, F: female, N: neutral, NB: non-binary, WF: weak female. UC: \textsc{uClassify}, RD: \textsc{Readable}, HF: \textsc{HackerFactor}, BT: \textsc{BERT}.}
	\label{tab1:samplecomments}
\vspace{-6mm}
\end{table}

\noindent \textbf{Audit of gender analyzers}:
In this work, we audit tools that predict the gender of the author from a piece of text. These are used for safety on social media~\cite{Na2011GenderIden} and, advertising~\cite{mukherjee2017gender}, but only work for the two binary genders - female and male, thus omitting non-binary authors.
Amazon's resume shortlister was scrapped\cite{reuters2018} as it was using gender-indicative words to shortlist only male candidates.
A more acute problem is when such analyzers are used for law enforcement operations like crime detection. AI tools are being used to determine user gender and link it with the content produced by them to understand how such content is expressed to aid crime detection and investigations. In cases of cybercrime, for instance, the identification of the gender from the content posted and linking these to the user's identity can narrow down the number of potential suspects~\cite{nonso}. However, since such AI tools are not trained to predict the non-binary class, such users are always going to be mispredicted thus placing them in a very risky position. 
Thus gender analyzers are important AI tools but come with their share of discrimination. Economical pricing, free-tier subscriptions, and easy to use web-interfaces have led to wide-scale adoption of these tools for both personal and professional use. One such tool, \textsc{Readable}, for instance, has Shopify, Netflix, Adobe, and NASA as clients. Here we audit~\cite{sandvig2014auditing} three very common existing tools that predict the gender of the author from a piece of text: two commercial tools -- \textsc{uClassify}~\cite{uclassify} and \textsc{Readable}~\cite{readable} and a free open-source tool -- \textsc{HackerFactor}~\cite{hackerfactor}. For a small subset of textual inputs, we also audit ChatGPT~\cite{chatgpt}, an LLM that generates human like text based on input prompts.

\noindent \textbf{Designing a fair gender analyzer}:
Next, we finetune a pre-trained BERT~\cite{devlin2018bert} based multi-label classifier which predicts three gender labels -- `male', `female', and `non-binary'. Our experiments using this off-the-shelf model show that while there are \textit{simple} ways to address existing societal biases, ML developers remain uninformed of these possible solutions.

In this study, we curate two datasets from the Reddit~\cite{reddit} (660k comments) and Tumblr~\cite{tumblr} (2.05M posts) platforms, from various self-identified male, female and non-binary individuals (more details in Section~\ref{sec:methodology}). The gender label for all posts on Tumblr and more than 20\% comments on Reddit are self-annotated by the author of the text. The rest have been annotated by two annotators based on the subreddits the post is collected from\footnote{Data available on request-- https://forms.gle/VVSrTxoYKPW16vCYA}. 
Some example comments by non-binary authors along with their associated predictions by the different platforms are available in Table~\ref{tab1:samplecomments}. The columns are labeled in the following order -- \textsc{uClassify}, \textsc{Readable}, \textsc{HackerFactor}, and our BERT model. We see that the BERT-based model correctly predicts the labels whereas the other platforms misclassify the statements.
Note that since the commercial classification models are black boxes and do not release the weights and training data, their outputs are not explainable. \textsc{HackerFactor}, though open-source, has fixed weights assigned to various \textit{gender-centric} words, which is used in a fixed formula to classify a given sentence. Finally, this investigation also provides us an opportunity to study various forms of machine learning biases~\cite{mehrabi2021survey} like \textit{representation, user interaction} and \textit{emergent bias}.

\noindent\textbf{Research questions}: Here, we state the key research questions that we address as part of this study.

\noindent \textbf{RQ1}: What are the gender predictions by the four platforms - \textsc{uClassify}, \textsc{Readable}, \textsc{HackerFactor} and BERT, for the comments from the male and female gender groups and whether the accuracy of the three existing tools is as high as they claim and how the finetuned BERT model performs in comparison to them. Through this question, we seek to verify if the tools are indeed accurate in identifying the gender of the binary authors through various standard measures like precision, recall, and F1 score.

\noindent\textbf{RQ2}: What is the predicted gender for the non-binary comments on all the platforms? Through this question, we attempt to gain insight into how these platforms perceive text written by non-binary authors and what kind of social implications this might have. Since the BERT model is trained to predict non-binary labels too, the accuracy values will give us an indication of how a gender-fair model may be designed and deployed.

\noindent\textbf{RQ3}: How do the audited platforms and more importantly, the BERT classifier, perform on the comments from two distinctly different social media platforms -- Reddit and Tumblr? Through this question, we attempt to understand the generalization capability of the BERT model for different textual datasets as well as compare the performance of the audited platforms on the two datasets.

\noindent \textbf{RQ4}: How do the advanced LLMs like \textsc{ChatGPT} perform when suitably prompted to classify gender based on the user posts.

\noindent\textbf{Our contributions}:
In this paper, we audit three existing gender analyzers that classify the gender of the author based on input text and also fine-tune and test a BERT-based multilabel classifier for Reddit and Tumblr comments from binary and non-binary authors. The existing systems -- \textsc{uClassify}, \textsc{Readable}, \textsc{HackerFactor} report an accuracy of $\approx 50\%$ on both Reddit and Tumblr for the binary authors' texts, lower than the minimum 70\% claimed by both \textsc{Readable} and \textsc{HackerFactor}. One of our BERT-based fine-tuned models has the highest overall accuracy of 83\% on Reddit and 66\% on Tumblr. For the non-binary authors' comments, on Tumblr, all the existing systems -- \textsc{uClassify}, \textsc{Readable} and \textsc{HackerFactor} predict female as the author's gender for a majority of the comments, whereas on Reddit, only \textsc{uClassify} and \textsc{Readable} report similar observations. Finally, in a transfer learning setup, our \textit{best} BERT-based model (fine-tuned on one dataset and then provided with few-shot examples from another dataset) has an accuracy of 90\% for non-binary authors, with an overall accuracy of 77\%, which demonstrates that even a simple off-the-shelf model can be accurately used to design a gender-inclusive system. Surprisingly, when we prompt \textsc{ChatGPT} in a zero-shot setting to perform gender classification based on the user comments from across the two datasets, the average accuracy is only 58\% with overall results being better for Reddit. This indicates that even such a powerful model is not very suitable as a general purpose gender analyzer.
\section{Related work}
On social media platforms, the task of gender identification is used for the purpose of security~\cite{fatemeh20}, advertising~\cite{mukherjee2017gender}, online safety~\cite{Na2011GenderIden} and opinion mining amongst others. Due to the different features of these social media websites, the task of gender identification is done based on the analysis of various user generated content like visual posts (display pictures~\cite{Vicente2019}), text (blog posts, comments, tweets~\cite{vashisth2020gender,sotelo2020gender,abdallah2020age, angeles2021gender, liu2018gender, christina2015gender}) etc. In this work, we focus on auditing commercial and open-source tools that are used for author gender identification through textual data on Reddit and Tumblr.

The early works in author gender identification were by Cheng et al.~\cite{Na2011GenderIden} to identify gender in short, content-free text, and, by Deitrick et al.~\cite{deitrick2012author} who used stylometric and word count features to classify email authors. Bartle and Zheng~\cite{bartle2015gender} proposed a Windowed  RCNN (WRCNN) that achieved an accuracy of 86\% on a blog dataset. Mukherjee and Bala~\cite{mukherjee2017gender} compared different machine learning models against commercial softwares for the binary gender task prediction and achieved a higher accuracy than the baseline. Vicente et al.~\cite{Vicente2019} used textual analysis of English and Portuguese tweets as part of their larger gender identification pipeline. Fatemeh et al.~\cite{fatemeh20} used an ANN based classifier coupled with whale optimization to identify the gender of email authors and report an accuracy of 98\%. Recently, there has been significant work in developing gender analyzers for English text~\cite{vashisth2020gender,sotelo2020gender,abdallah2020age}. Vasilev's \cite{vasilev2018inferring} thesis deals with inferring gender of Reddit users, but like all existing works in this domain, including ones described above, it treats gender as a binary attribute.

It is apparent that there has been considerable research in this domain but none of them have focused on addressing the problem for non-binary gender groups, thereby allowing the existing discrimination to fester. We also notice that while all of the previous works have studied design and development of open-source models or compared against commercial products, none of them have focused on auditing any of these models. We address both these problems in this paper by auditing commercial and open-source third-party softwares for multiple gender groups, and fine-tuning a pretrained model that is able to make predictions for the non-binary gender, thus presenting a gender-inclusive alternative to the existing softwares.
\section{Dataset \& Methodology}
\label{sec:methodology}

\subsection{Dataset curation}

In this study, we create social media comments datasets from two platforms-- Reddit and Tumblr, in the English language. 
\begin{compactitem}
    \item \textsc{Reddit}: Top-level comments are collected from various subreddits belonging to male, female and non-binary interests. We assign a subreddit as \textbf{X}-interest (where \textbf{X} is either male, female or non-binary) if the subreddit name or description refers to any of the corresponding gender categories.
    \item \textsc{Tumblr}: Blogs (unique per user) are collected and segregated based on the presence of the gender label in the blog bio. This is followed by the collection of self-posts and answers given by the bloggers, per blog. Each such post or answer (referred to as comment henceforth) is assigned an \textbf{X}-gender (where \textbf{X} is either male, female, or non-binary) based on the pronoun or gender mentioned by the blogger in their profile description.
\end{compactitem}

More details from the two datasets are present in Table~\ref{tab:datasetinfo}

\noindent \textbf{Gender annotation strategy}: For Reddit, over 20\% of the data is self-annotated for gender by the commenters themselves. We annotate the rest
as either male, female, or non-binary based on the subreddit it is collected from. We believe our annotation strategy is effective because all of these subreddits serve the interests of the corresponding gender group exclusively and thus the posts and comments are only by individuals who identify with that particular gender\footnote{https://tinyurl.com/4mkkud2j}. On Tumblr, all posts are self annotated by the bloggers.

\begin{table}[!t]
    \tiny
    \centering
    \begin{tabular}{| l | c | c | c | c | c | c || c | c | c | c | c |}
    \hline
    & \multicolumn{6}{c||}{\textbf{Reddit}} & \multicolumn{5}{c|}{\textbf{Tumblr}}\\
    \hline
    \textbf{G} & \textbf{\#S} & \textbf{\#C} & \textbf{$L_{avg}$} & \textbf{\# $A_{unq}$} & \textbf{\# $P_{unq}$} & \textbf{\# $(C/P)_{\mu}$} & \textbf{\#B} & \textbf{\#C} & \textbf{$L_{\mu}$} & \textbf{\# $A_{unq}$} & \textbf{\# $(C/B)_{\mu}$}  \\
    \hline
	M & 3 & 240k & 99 & 62k & 33k & 7 & 230 & 343k & 99 & 230 & 1491\\
	\hline
	F & 2 & 240k & 99 & 67k & 29k & 8 & 688 & 704k & 141 & 688 & 1023\\
	\hline
    NB & 7 & 180k & 80 &44k & 65k & 3 & 1670 & 1.01M & 132 & 1670 & 605\\
	\hline\hline
	T & 12 & 660k & 93 & 173k & 127k & 6 & 2588 & 2.05M & 124 & 2588 & 1040\\
	\hline 
    \end{tabular}
    \vspace{-2mm}
    \caption{\footnotesize Details of the subreddits/blogs, comments, authors and posts in the Reddit \& Tumblr comment dataset. On Tumblr, each blog post is available with a unique UUID, without an author username, hence the number of unique authors is the same as the number of unique blogs. For Reddit, \#S is subreddit count and \#$P_{unq}$ is count of unique posts. On both platforms, \#C is comment count, $L_{\mu}$ is the avg. length of a comment, \# $A_{unq}$ is count of unique authors and \# $(C/P)_{\mu}$ ($(C/B)_{\mu}$) is the avg. number of comments per post (blog). B: blog, M: male, F: female, NB: non-binary.}
    \label{tab:datasetinfo}
\end{table}

Going by this annotation scheme, we note a few more interesting details about the dataset in Table \ref{tab:datasetinfo}. The Reddit dataset contains approximately 173k unique authors, of which male and female comments are made by about $\approx 65k$ unique authors each and non-binary comments are by $\approx 44k$ unique authors. Female and male subreddits have about 32k unique posts whereas non-binary subreddits have $\approx 65k$ unique posts. Finally, male and female posts have an average of 8 comments per post where as non-binary posts have an average of 3 comments per post. Similarly for Tumblr, the dataset contains 2588 unique bloggers, of which there are 230 male, 688 female, and 1670 non-binary bloggers. Male and female blogs have more than 1K posts per blog whereas non-binary blogs have $\approx 600$ posts per blog.

\subsection{Methodology}
\noindent \textit{Post collection}: We collect the posts/blogs from both platforms using \textsc{Pushshift IO}\footnote{https://api.pushshift.io} and \textsc{Tumblr API\footnote{tumblr.com/docs/en/api/v2}} respectively. The data for Reddit is collected for posts between Jan 2017 and Aug 2022, and for Tumblr, blogs posted between May 2008 and Nov 2022. On Reddit, we collect only the top-level comments for each post. Similarly, for Tumblr, we collect all the posts by the blogs filtered for the gender group. The three platforms are now described in brief, followed by a short description of the BERT based model. 
 
\noindent {\bf \textsc{uClassify}:} This is a ML web service that provides over 120 publicly available text-based classifiers that perform tasks like sentiment, gender, age and even Myer-Briggs analysis for a given piece of text. The service is accessible through a web interface, REST APIs and SDKs with multiple programming language bindings and for multiple languages like English, Spanish, French and Swedish. 

\noindent{\bf \textsc{Readable}:} This is a platform that scores textual data to improve readability and utilizes multiple formulae like Gunning Fog index and SMOG index to evaluate the input text. The set of free classifiers available on the website can be used for gender analysis, profanity and buzzword detection, etc. The service is accessible either through a web interface or REST API for the English language. 

\noindent{\bf \textsc{HackerFactor}:} This platform offers an open source software called Gender Guesser that determines an author's gender by using vocabulary statistics. While the software is not ML based, it is highly popular and freely available to use through a web interface. Users may also copy the code and run it locally. 

\noindent{\bf \textsc{BERT} base model:} We finetune a pretrained BERT base-uncased model to predict three classes -- male, female and non-binary. This is a transformer model that has been pretrained on a large corpus of English data. We finetune the model for multiple settings on both datasets. On both datasets, the train-validation-test split of the comments per gender group is $70:10:20$. The batch size was set to $64$, learning rate to $2.0e^{-5}$, seed value to $4$, dropout to $0.1$ and optimizer to \textsc{ADAM}. 

\noindent{\bf \textsc{ChatGPT}:} This is a large language model that generates human-like text based on prompts. We prepare appropriate prompts that allow it to simulate a text-based gender analyzer and predict the gender for the input text's author.

\subsection{Steps for experimentation}

The steps for experimentation on the three platforms are described in brief as follows -
\begin{compactitem}
    \item \textsc{uClassify}: A Python script issues POST requests to the \textbf{genderanalyzer\textunderscore v5} REST API endpoint for every comment and stores the responses (either male or female), which are then finally analyzed against the ground truth data to calculate the metrics.
    \item \textsc{Readable}: The comments are submitted to the web interface of the \textit{gender analyzer} tool through Selenium web automation which also collects the responses (either male, female or neutral) and finally calculates the accuracy metrics.
    \item \textsc{HackerFactor}: All the comments are processed locally by the freely available source code downloaded from the Gender Guesser tool of \textsc{HackerFactor}. The code evaluates the text for both formal and informal types, and we capture the predicted gender for the informal type.
    \item \textsc{ChatGPT}: Due to budget constraints, only 1500 comments (500 from each gender group) per platform are audited here. We generate appropriate prompts that simulate a text-based gender analyzer.
\end{compactitem}

For the \textsc{BERT} base model, we experiment with multiple finetuning settings using the datasets from Reddit ($\mathbb{R}$) and Tumblr ($\mathbb{T}$) to address our research questions from Section~\ref{sec:intro}.
\begin{compactitem}
    \item Finetuning with comments from $\mathbb{R}$ and $\mathbb{T}$ individually.
    \item Finetuning by combining equal percentage of comments from $\mathbb{R}$ and $\mathbb{T}$.
    \item Finetuning first with comments from one platform and then again finetuning for a second time in a few-shot setting using comments from the other platform. 
\end{compactitem}

\section{Observations \& Results} 

\begin{table}[!t]
    \tiny
	\begin{center}
		\begin{tabular}{ | c | c| c | c |}
            \hline
			\textbf{Platform} & \textbf{Training data} & \textbf{Claimed Acc.} & \textbf{Predicted classes} \\
			\hline
			\textsc{uClassify} & Blogs & N.A. & M, F\\
			\hline
			\textsc{Readable} & N.A & 70\% & M, F, N \\
			\hline
			\textsc{HackerFactor} & N.A. & 70\%  & \makecell[c]{M, F, UNK, \\ WM, WF, WU} \\
			\hline
			\textsc{BERT} base & WIKI \& book-corpus & N.A. & M, F, NB \\
			\hline
		\end{tabular}
	\end{center}
	\vspace{-4mm}
 \caption{\footnotesize Training dataset and claimed accuracy for the four platforms. The BERT model is fine-tuned on the two comment datasets for evaluation. M: male, F: female, N: neutral, NB: non-binary, UNK: unknown, WM: weak male, WF: weak female, WU: weak unknown.}
	\label{tab:models}
\end{table}

\subsection{Performance on male and female comments (RQ1, RQ3)}

We first discuss the results for the three baseline platforms --\textsc{uClassify}, \textsc{Readable} and \textsc{HackerFactor} for the cisgender binary comments. \textsc{uClassify} is trained on a balanced dataset of 11,000 blogs (divided equally between males and females), but no baseline accuracy is provided. \textsc{Readable} and \textsc{HackerFactor} platforms mention a baseline accuracy of $70\%$ for the gender analyzer but do not provide any information on the training data or previous evaluation metrics. More information regarding the classes predicted by each platform is present in Table~\ref{tab:models}.


\begin{figure*}[t!]
\begin{minipage}[]{0.5\linewidth}
	\scriptsize
		\centering
		\begin{tabular}{ | c | c| c| c|| c| c || c | c || c | c | } 
            \hline
            \multirow{2}{*}{\textbf{Platform}} & \multirow{2}{*}{\textbf{Gender}} & \multicolumn{2}{c||}{\textbf{Accuracy}} & \multicolumn{2}{c||}{\textbf{Precision}} & \multicolumn{2}{c||}{\textbf{Recall}} & \multicolumn{2}{c|}{\textbf{F1}}\\
            \cline{3-10}
		      & & \textbf{$\mathbb{R}$} & \textbf{$\mathbb{T}$} & \textbf{$\mathbb{R}$} & \textbf{$\mathbb{T}$} & \textbf{$\mathbb{R}$} & \textbf{$\mathbb{T}$} & \textbf{$\mathbb{R}$} & \textbf{$\mathbb{T}$}\\
			\hline
			\multirow{3}{*}{\textsc{UCLF}} & Overall &  $54.2\%$ & $52.2\%$ & $63\%$ & $54.3\%$ & $54.2\%$ & $52.2\%$ & $45.2\%$ & $49.8\%$\\
			\cline{2-10}
			 & Male & $13.4\%$ & $15.5\%$ & $73.5\%$ & $40.3\%$ & $13.4\%$ & $15.5\%$ & $22.8\%$ & $22.4$\%\\ 
			\cline{2-10}
		     & Female & $95.1\%$ & $88.8\%$ & $52.4\%$ & $68.3\%$ & $95.1\%$ & $88.8\%$ & $67.6\%$ & $77.2\%$\\
			\hline \hline
			\multirow{3}{*}{\textsc{RDBL}} & Overall &  $47.8\%$  & $41.6\%$ & $56.7\%$ & $50\%$ & $47.8\%$ & $41.6\%$ & $51.5\%$ & $44.8\%$\\
			\cline{2-10}
			 & Male &  $41.0\%$ & $36.9\%$ & $57.5\%$ & $32.9\%$ & $41.0\%$ & $36.9\%$ & $47.8\%$ & $34.8\%$\\
			\cline{2-10}
		     & Female &  $54.7\%$ & $46.2\%$ & $55.9\%$ & $67.1\%$ & $54.7\%$ & $46.2\%$ & $55.3\%$ & $54.7\%$\\
			\hline \hline
			\multirow{3}{*}{\textsc{HCFTR}} & Overall &  $56.8\%$ & $48.6\%$ & $57.6\%$ & $49.4\%$ & $56.8\%$ & $48.6\%$ & $57.2\%$ & $48.3\%$\\
			\cline{2-10}
			 & Male &  $55.2\%$ & $42.8\%$ & $57.9\%$ & $32.1\%$ & $55.2\%$ & $42.8\%$ & $56.5\%$ & $36.7\%$\\
			\cline{2-10}
		     & Female &  $58.5\%$ & $54.4\%$ & $57.3\%$ & $66.6\%$ & $58.5\%$ & $54.4\%$ & $57.0\%$ & $59.9\%$\\
			\hline
		\end{tabular}
		\vspace{-2 mm}
\end{minipage} \hfill
\begin{minipage}[]{0.34\linewidth}
\centering
        \includegraphics[height=3.15cm, keepaspectratio]{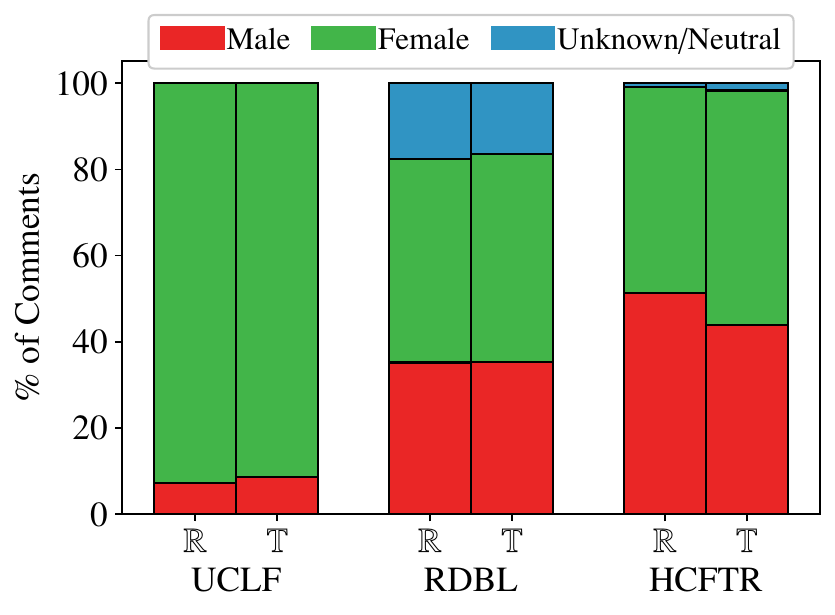}
        \vspace{-2mm}
\end{minipage}
\captionsetup{width=0.99\textwidth}
\captionof{figure}{\footnotesize Accuracy of \textsc{uClassify} (UCLF), \textsc{Readable} (RDBL) and \textsc{HackerFactor} (HCFTR) on comments by male, female (in Table) \& non-binary (in Figure) authors from the two datasets -- Reddit ($\mathbb{R}$) and Tumblr ($\mathbb{T}$). From the table, we see the platforms have low accuracy, independent of the dataset. Male comments have lower accuracy than females on all platforms. From the figure, we see all platforms predict female as the gender most often for non-binary authors, independent of the dataset. \textsc{Readable} is the most `fair' platform.}
\vspace{-5mm}
\label{fig:existing_plat_accuracy}
\end{figure*}

\noindent \textit{Observations.} The table in Figure~\ref{fig:existing_plat_accuracy} presents the results for the accuracy of prediction for comments by male and female individuals on the \textsc{Reddit} (480k comments) and \textsc{Tumblr} datasets (1.04M comments). \textsc{Readable} reports the lowest accuracy for both datasets -- 47.8\% and 41.6\% respectively. The best performing platform for \textsc{Reddit} is \textsc{HackerFactor} ($\approx 57\%$) and for \textsc{Tumblr} is \textsc{uClassify} (52\%). The highest precision on both datasets is reported by \textsc{uClassify} -- 63\% and 54.3\% respectively. For recall and F1, we note that \textsc{uClassify} is the best for \textsc{Tumblr} and \textsc{HackerFactor} is the best for \textsc{Reddit}.

The accuracy for male comments is lower than for females on all platforms. The lowest value is reported for \textsc{uClassify} -- $13.4\%$ and $15.5\%$ respectively and the highest on \textsc{HackerFactor} -- $55\%$ and $43\%$ respectively. Thus, it is clear that the existing platforms are not very accurate in predicting the gender of comments by male authors. Similar observations can be made regarding the recall and F1. Thus, the existing models are overcompensating and misclassifying the male comments as female. It is difficult to know whether this is due to the training data, the weights of the gendered words, or the model inference mechanism. Interestingly, the precision for males is higher than for females for Reddit comments. There is a $> 70\%$ gap between the female and male accuracies on \textsc{uClassify}, independent of the dataset while this difference is less stark on the other platforms. The accuracy for the non-binary gender group is not shown as none of these platforms predict a non-binary gender.

\subsection{Performance on non-binary comments (RQ2, RQ3)}
Next, we study the responses of the three platforms for the non-binary dataset. Since these platforms do not predict a non-binary class, we cannot measure the accuracy. We thus study only the predictions for all comments on these platforms. 

\noindent\textit{Observations.} In the bar plots in Figure~\ref{fig:existing_plat_accuracy}, we show the distribution of predictions for the non-binary author comments by the three gender analyzer platforms. We can see that for \textsc{uClassify} ($> 91\%$) and \textsc{Readable} ($> 47\%$), the majority prediction is female, independent of the dataset. On \textsc{HackerFactor}, the majority prediction for Reddit is male ($\approx 51\%$), but on Tumblr, the majority is again female ($\approx 54\%$). On \textsc{Readable}, $\approx 18\%$ of Reddit comments and $16.5\%$ Tumblr comments are predicted as neutral. \textsc{HackerFactor} predicts the `unknown' label for less than 2\% of the inputs.

\subsection{Performance of the BERT classifier (RQ1, RQ2, RQ3)}

In this final segment, we test a BERT-base classifier under different fine-tuning settings described in Section~\ref{sec:methodology} to verify if a simple classifier, finetuned on social media data from our two datasets can be used to mitigate some of the existing biases observed in the audit above. For these experiments, we choose 660k comments from each dataset (240k comments from male and female authors each, and 180k comments from non-binary authors) unless otherwise stated. 

\begin{figure*}[!t]
	\centering
	\begin{subfigure}{0.32\textwidth}
	\centering
		\includegraphics[height=2.5cm, keepaspectratio]{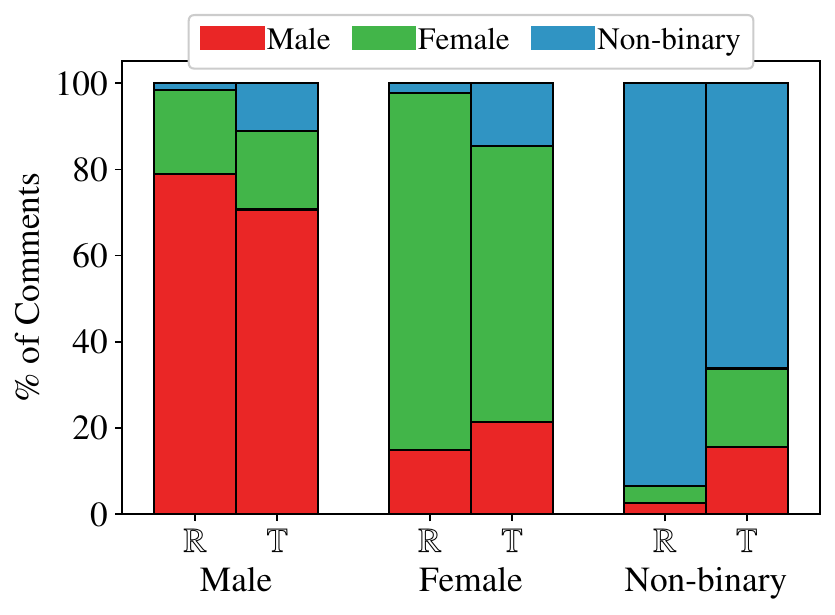}
		\vspace{-2mm}\caption{\footnotesize Fine-tune and test on the same dataset.}
		\label{fig:bert_fine_tuned_same_tested_same}
	\end{subfigure}%
 \vspace{-3mm}
	~\begin{subfigure}{0.32\textwidth}
    \centering		
		\includegraphics[height=2.5cm, keepaspectratio]{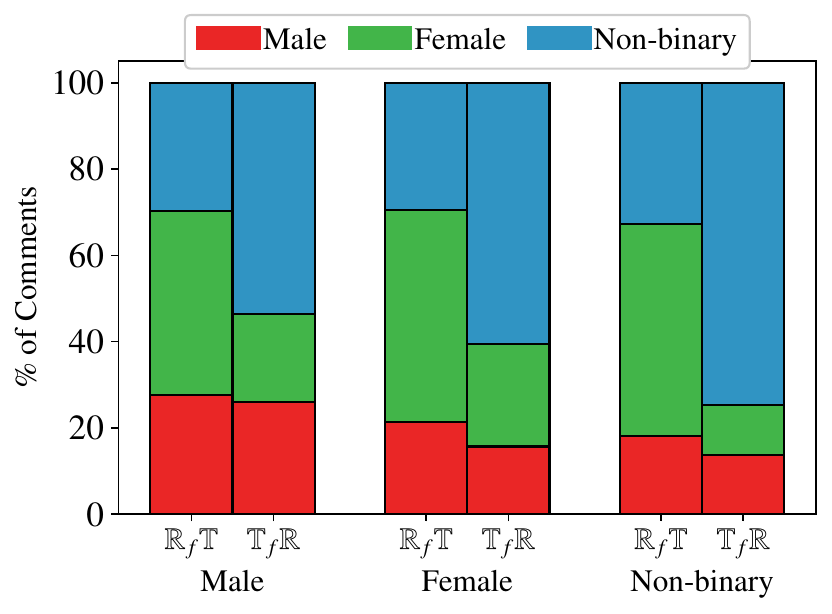}
		\vspace{-2mm}\caption{\footnotesize Fine-tune and test on different datasets.}
		\label{fig:bert_fine_tuned_same_tested_diff}
	\end{subfigure} 
	\begin{subfigure}{0.32\textwidth}
	\centering
		\includegraphics[height=2.5cm, keepaspectratio]{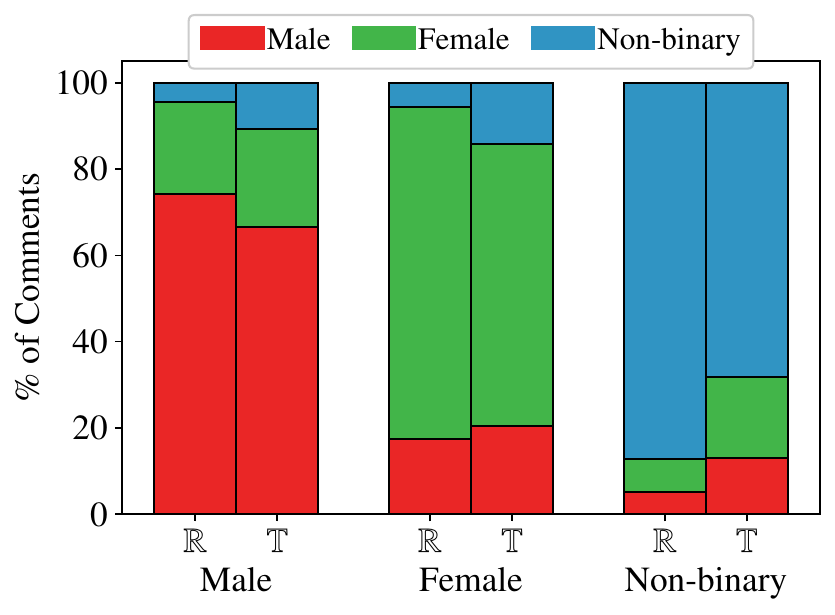}
		\vspace{-2mm}\caption{\footnotesize Fine-tune by mixing the two datasets equally.}
		\label{fig:bert_fine_tuned_mix_tested_mix}
	\end{subfigure}%
	\caption{\footnotesize Accuracy for the BERT model under different fine-tuning settings when testing on the Reddit ($\mathbb{R}$) and Tumblr ($\mathbb{T}$) datasets. $\mathbb{R}_f\mathbb{T}$ refers to the model fine-tuned on Reddit and tested on Tumblr and $\mathbb{T}_f\mathbb{R}$ refers to the model fine-tuned on Tumblr and tested on Reddit. Each bar shows how many comments for each gender group are classified as `male', `female', and `non-binary'. }
	\label{fig:bert_fine_tuning}
	\vspace{-4mm}
\end{figure*}

\subsubsection{\textbf{Fine-tuning entirely on one dataset}}
In this experiment, we check the model performance by fine-tuning entirely on comments from one dataset and then testing on comments from the same dataset (as baseline) and the other dataset (for generalizability). Figure~\ref{fig:bert_fine_tuned_same_tested_same} presents the results for the first scenario and Figure~\ref{fig:bert_fine_tuned_same_tested_diff} presents the results for the second scenario. 

\noindent\textit{Observations.} From Figure~\ref{fig:bert_fine_tuned_same_tested_same} we note that the model performs very well on the comments from Reddit, with the overall accuracy being 84.4\%. In contrast, the overall accuracy for Tumblr is only 67.1\%. This shows that the model learns better from the comments on Reddit than on Tumblr. On looking at the results for each gender group, it is evident that the BERT model can identify non-binary authors' comments particularly well in the Reddit dataset, but the accuracy for cisgender binary authors, especially males (79\%) is comparatively low. On the Tumblr dataset, we observe that the best performance is reported for males ($\approx 71\%$), and only 66\% of the non-binary authors' comments are classified correctly. Compared to the existing platforms -- \textsc{uClassify}, \textsc{Readable} and \textsc{HackerFactor}, the BERT models report higher accuracy, on both datasets. This shows that even a simple baseline model can not only improve performance (for cisgender binary authors) but also be fair toward sensitive groups (non-binary authors) within the target population.

Next, in Figure~\ref{fig:bert_fine_tuned_same_tested_diff}, we evaluate the generalizability across datasets by testing a Reddit fine-tuned model on the Tumblr dataset ($\mathbb{R}_f\mathbb{T}$) and vice-versa ($\mathbb{T}_f\mathbb{R}$). We see a drastic drop in performance -- accuracies of $37\%$ for $\mathbb{R}_f\mathbb{T}$ and $\approx 38\%$ for $\mathbb{T}_f\mathbb{R}$. This indicates that there is a stark difference in the language dynamics of the two datasets and the model cannot generalize the learning from one to another. Interestingly, while in $\mathbb{R}_f\mathbb{T}$, the highest accuracy is reported for females -- 49\%, non-binary authors' accuracy is 32.8\%. On the other hand, in $\mathbb{T}_f\mathbb{R}$ we see that the lowest accuracy is reported for female authors -- $\approx 24\%$ and the accuracy for non-binary authors is 75\%. Hence, a model trained on Tumblr can identify comments from non-binary authors on Reddit better than on Tumblr itself. This is possibly because the model trained on Tumblr sees more nuanced examples thus constructing a more complex decision boundary that is able to efficiently classify the non-binary comments from Reddit that are typically more distinct in nature.

\subsubsection{\textbf{Fine-tuning using a mixed dataset}}
To understand how a generalizable model may be designed, we fine-tune BERT using comments from both Reddit and Tumblr. We randomly sample 50\% comments for each gender group from the two datasets to create a mixed fine-tuning dataset. 

\noindent\textit{Observations.} In Figure~\ref{fig:bert_fine_tuned_mix_tested_mix}, we report the results for this experiment. We can immediately observe gains from this setup as the overall accuracies improve to 79\% ($\mathbb{R}$) and 67\% ($\mathbb{T}$) compared to the case of fine-tuning on one and testing on the other dataset. For individual gender groups, we see that this mixed setup performs poorly on the Reddit dataset as compared to the original setup where the model was trained completely on Reddit. All gender groups report a drop in accuracy by $\approx 5\%$. For Tumblr, the results are different -- only the accuracy of male authors reduces by 4\%, with there being a slight improvement in accuracy for both female authors (1.3\%) and non-binary authors (2\%). Thus, while such a setup does not give any significant improvement for Tumblr and in fact reports lower accuracy for Reddit, it is more generalizable than fine-tuning on only one dataset. For the cisgender binary groups, this model is better than both \textsc{Readable} and \textsc{HackerFactor}, and for the male authors on \textsc{uClassify}.

\subsubsection{\textbf{Few-shot learning with limited examples}} The final setup we test for is also the most realistic scenario that is observed in real-world social networks. It is possible that none of the existing gender analyzers have trained their models on non-binary data because large-scale gender-labeled data is extremely difficult to collect, especially for the non-binary gender group. Thus, we envision a scenario where the BERT model has been fine-tuned on data from one dataset (for example -- Reddit or Tumblr in our work) and is available for use by other platforms. The other platforms may have a very small number of labeled examples of all gender groups available through voluntary self-disclosure which can then be used to perform a second round of fine-tuning, similar to a few-shot learning scenario, and the model can then be deployed for use on the said platform. 

\begin{figure*}[!t]
	\centering
	\begin{subfigure}{0.32\textwidth}
	\centering
		\includegraphics[height=2.5cm, keepaspectratio]{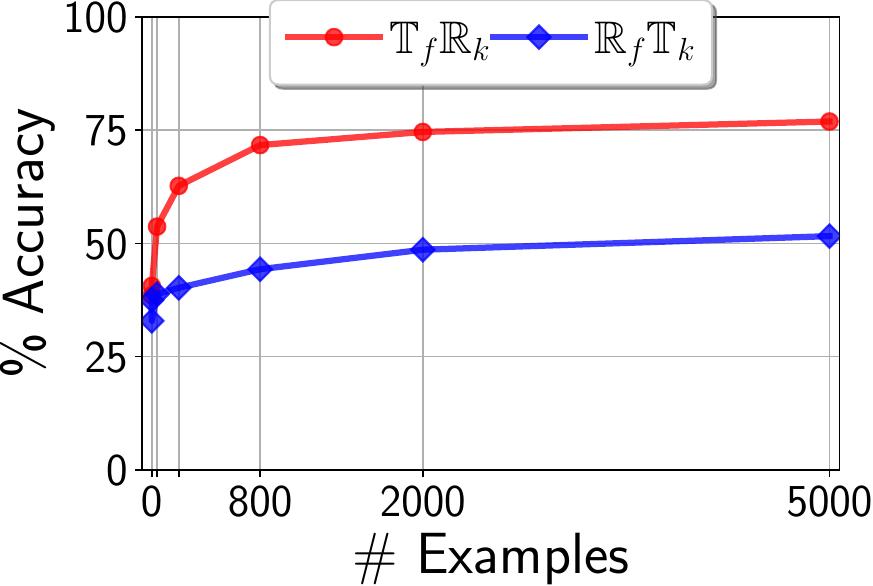}
		\vspace{-2mm}\caption{\footnotesize Overall accuracy for $\mathbb{T}_f\mathbb{R}_k$ \& $\mathbb{R}_f\mathbb{T}_k$} 
		\label{fig:bert_kshot_overall}
	\end{subfigure}%
 \vspace{-3mm}
	~\begin{subfigure}{0.32\textwidth}
    \centering		
		\includegraphics[height=2.5cm, keepaspectratio]{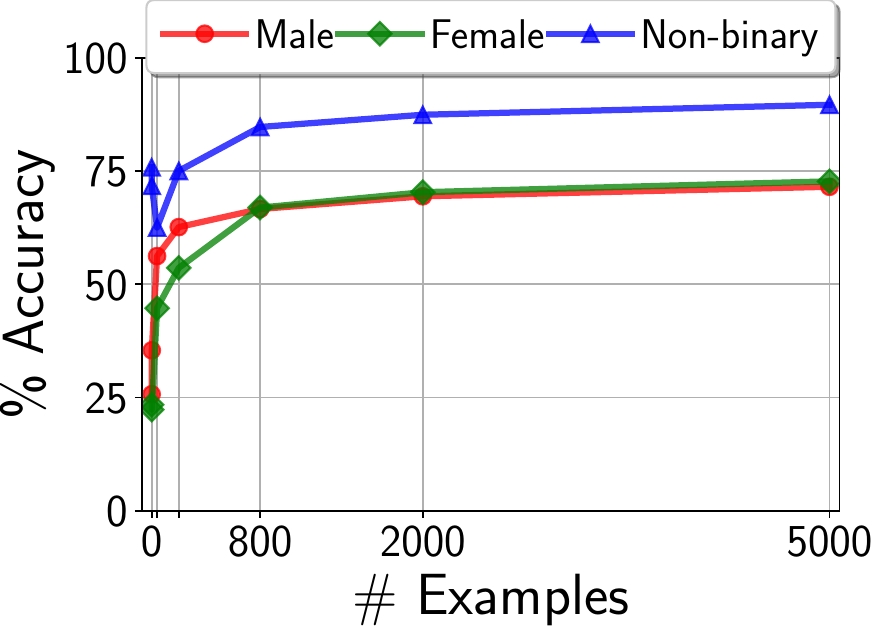}
		\vspace{-2mm}\caption{\footnotesize Per gender accuracy for the $\mathbb{T}_f\mathbb{R}_k$ setup.}
		\label{fig:bert_kshot_tfrx}
	\end{subfigure} 
	\begin{subfigure}{0.32\textwidth}
	\centering
		\includegraphics[height=2.5cm, keepaspectratio]{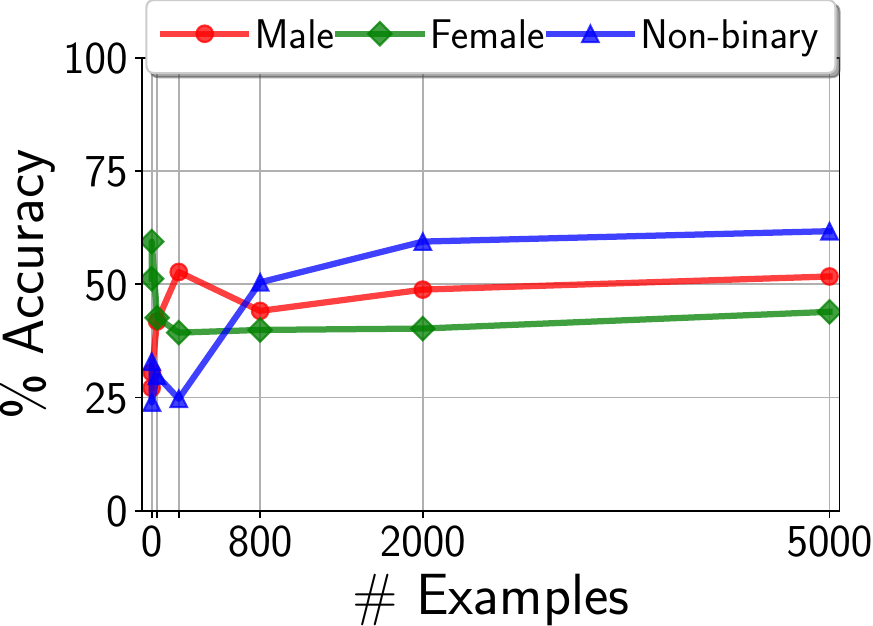}
        \vspace{-2mm}
		\caption{\footnotesize Per gender accuracy for $\mathbb{R}_f\mathbb{T}_k$ setup.}
		\label{fig:bert_kshot_rftx}
	\end{subfigure}%
	\caption{\footnotesize Accuracy for the few-shot learning scenario. Two models are evaluated-- fine-tuning on Tumblr with few-shot examples from Reddit ($\mathbb{T}_f\mathbb{R}_k$) and vice versa ($\mathbb{R}_f\mathbb{T}_k$).~$k$ signifies the number of examples under consideration, ranging from 0 to 5000. ($\mathbb{T}_f\mathbb{R}_k$) has the overall best performance as well as the best performance for non-binary data with 5000 examples from Reddit.}
	\label{fig:bert_kshot}
	\vspace{-6mm}
\end{figure*}

We experimentally verify this scenario for the following values of $k$ (number of shots) -- $\{0, 1, 40, 200, 800, 2000, 5000\}$. The model is first fine-tuned using the entire training data (70\% of the total) from one of the platforms. This is followed by a $k$-shot learning and testing using data from the second dataset. Thus, $\mathbb{T}_f\mathbb{R}_{200}$ implies that the model is first fine-tuned using Tumblr's data and then 200 examples from each gender group in Reddit's dataset are used to perform a second round of learning. The results are noted in Figure~\ref{fig:bert_kshot}. We perform 5-fold cross validation and report the average accuracy. 

\noindent\textit{Observations.} From Figure~\ref{fig:bert_kshot_overall} we see that the model fine-tuned on Tumblr and provided with $k$ examples from Reddit always performs better than the opposite scenario (average accuracy difference of 18\%). In the 0- and 1-shot setting, the accuracy of both models is low -- 38.5\% and 37.5\% for $\mathbb{T}_f\mathbb{R}_0$ \& $\mathbb{R}_f\mathbb{T}_0$ respectively and, 40.6\% and 33\% for $\mathbb{T}_f\mathbb{R}_1$ \& $\mathbb{R}_f\mathbb{T}_1$ respectively. For $\mathbb{T}_f\mathbb{R}_{5000}$ \& $\mathbb{R}_f\mathbb{T}_{5000}$, the overall accuracies are 77\% and 52\% respectively. Thus it is clear that $\mathbb{T}_f\mathbb{R}_{k}$ is the better model and generalizes well with very few examples from Reddit.

On looking at the gender group-wise accuracy for both models in Figures~\ref{fig:bert_kshot_tfrx} and ~\ref{fig:bert_kshot_rftx}, we see that with an increasing number of examples, the accuracy for non-binary gender group is higher than the cisgender group. From Fig.~\ref{fig:bert_kshot_tfrx}, we see that $\mathbb{T}_f\mathbb{R}_{k}$ has more than 40\% difference in accuracy between the non-binary and cisgender binary authors for the 0- vs 1-shot learning scenario. This difference is 25\% on average, across all $k$-shots. The model achieves 90\% accuracy for non-binary authors and 72\% for the cisgender binary authors after training with only 5000 examples from Reddit. For $\mathbb{R}_f\mathbb{T}_{k}$ (Fig.~\ref{fig:bert_kshot_rftx}), we see that not only is the overall accuracy low, but the gender-wise accuracies are also low. While the accuracies do increase with an increasing number of examples, the growth is not very significant. The maximum accuracy for males and non-binary authors is observed for $\mathbb{R}_f\mathbb{T}_{5000}$ -- 51.7\% for males and 61.7\% for non-binary. The accuracy for females is 42\%, with the highest being observed in the 1-shot scenario at 59.4\%.

These results indicate that few-shot learning works when only a limited set of examples is available and the generalizability from Tumblr to Reddit is higher than from Reddit to Tumblr. 

\noindent \textit{Overall takeaways.} The following conclusions can be drawn from the above subsections.
\begin{compactitem}
\item The prediction accuracy for all existing platforms is $\approx 53\%$ for comments from Reddit and $\approx 47.5\%$ for Tumblr. Thus, the platforms perform poorly on both datasets, with an accuracy that is worse than a coin-toss for Tumblr. The accuracy is higher for females on all platforms. This indicates possible \textit{representation} bias, whereby the sampling of the training dataset on the platforms may have idiosyncrasies.
\item None of the platforms show good performance on either dataset, indicating a lack of generalizability.
\item The commercial third-party platforms predict the majority of the non-binary comments to be female, with \textsc{uClassify}'s predictions being overwhelmingly female. \textsc{HackerFactor} predicts majority female for Reddit and majority male for Tumblr. As noted earlier, if cybercrime investigations would resort to such tools for gender identification then most of the non-binary individuals would be at risk of gender misprediction and may be exposed to unnecessary harassment given that there is a recent surge in the number of cybercrimes committed by women\footnote{\url{https://www.trendmicro.com/vinfo/us/security/news/cybercrime-and-digital-threats/gender-in-cybercrime}}.
\item While both \textsc{Readable} and \textsc{HackerFactor} have a third prediction class -- `neutral' and `unknown' respectively, only \textsc{Readable} has a non-negligible prediction for this class. Hence, it is slightly fairer among the two platforms. 
\item A fine-tuned BERT-based multilabel classifier works better than all the existing platforms that we have audited for the task of gender classification. It can be trained to predict non-binary gender labels as well, with sufficiently high accuracy.
\item A BERT model fine-tuned on a mixed dataset from both Reddit and Tumblr generalizes better than the one fine-tuned on only one dataset. It should be noted that such a model is still comparable to the few-shot setting.
\end{compactitem}

\subsection{Performance comparison with \textsc{ChatGPT} (RQ4)}

\begin{table}[!t]
    \centering
    \scriptsize
    \begin{tabular}{|c|c|c|c|c||c|c|c|c|}
    \hline
        \multirow{2}{*}{\textbf{Platforms}} & \multicolumn{4}{c||}{$\mathbb{R}$}  & \multicolumn{4}{c|}{\textbf{$\mathbb{T}$}} \\ 
        \cline{2-9}
         & \textbf{M} & \textbf{F} & \textbf{NB} & \textbf{Overall} & \textbf{M} & \textbf{F} & \textbf{NB} & \textbf{Overall} \\ \hline\hline
        \textbf{\textsc{ChatGPT}} & \textbf{80.8} & 58.4 & 87.8 & 75.7 & 26.2 & 44.6 & \textbf{48.2} & 39.7 \\ \hline
        \textbf{BERT ($\mathbb{T}_f\mathbb{R}_{5000}$)} & 75.8 & 73 & \textbf{88.4} & \textbf{79.1} & \textbf{55.8} & 60.8 & 40.6 & \textbf{52.4} \\ \hline
        \textbf{\textsc{uClassify}} & 14.8 & \textbf{95.4} & X & 55.1 & 16.8 & \textbf{88.8} & X & 35.2 \\ \hline
        \textbf{\textsc{Readable}} & 41 & 51.4 & X & 30.8 & 30.8 & 48.4 & X & 26.4 \\ \hline
        \textbf{\textsc{HackerFactor}} & 55.4 & 57.2 & X & 56.3 & 43.8 & 57.8 & X & 50.8 \\ \hline
    \end{tabular}
    \vspace{-2mm}
    \caption{\footnotesize Accuracy on 1500 comments (500 from each gender group) from each dataset for all platforms. On the Reddit dataset, \textsc{ChatGPT}'s performance is comparable to our best finetuned BERT model and far superior to the other platforms. On the Tumblr dataset, \textsc{ChatGPT}'s performance is best for the non-binary comments. For male and female comments, \textsc{ChatGPT} performs worse than the other platforms. Maximum values are in bold. M: male, F: female, NB: non-binary.}
    \label{tab:chatgpt}
        \vspace{-4mm}
\end{table}

LLMs are being adopted on a wide-scale for general purpose and domain specific tasks these days. We simulate a text-based gender analyzer on \textsc{ChatGPT}, one of the most sophisticated LLMs available presently and audit it for a small sample of comments (500 from each gender group) from the two datasets. This allows us to understand its performance in a zero-shot scenario for a domain-specific task. We also compare it against the other available tools as well as our best finetuned model. The results are presented in Table~\ref{tab:chatgpt}. We see that its performance is comparable to our BERT model for the Reddit dataset, with the best performance reported for male comments and almost as good as the BERT model for non-binary comments. Conversely, on the Tumblr dataset it performs worse than all other platforms for the male/female comments but performs better than BERT for the non-binary comments. Overall, even such a powerful model is not able to report a sustained performance across all the three classes and the two datasets (albeit a small sample due to budget constraints) and therefore one needs to use it with appropriate caution (and possibly after improvement) for the simulation of the gender analysis task.

\section{Discussion and conclusion}
In this paper, we audit various gender analyzer platforms viz. \textsc{uClassify}, \textsc{Readable} and \textsc{HackerFactor}. Overall results indicate a prediction accuracy of less than 60\% on both the Reddit and Tumblr datasets. The platforms perform better for text by female authors than by male authors, especially on \textsc{uClassify}. Thus, both \textsc{uClassify} and \textsc{Readable} seem to be over-compensating and are predisposed to predict female for the input data. \textsc{HackerFactor} has a more balanced output but is still not very accurate for either gender. Thus to answer \textbf{RQ1} -- the existing tools are not accurate at identifying the gender of binary authors, reporting lower accuracy than what they claim. The performance on the two datasets (from the table in Figure~\ref{fig:existing_plat_accuracy}) answers \textbf{RQ3} -- the reported accuracies are similar, albeit poor.

Next, we audit these tools for non-binary authors' text and observe that all platforms are \textit{female-leaning}, thus reinforcing the societal belief about non-binary individuals being considered more effeminate. \textsc{uClassify} doesn't predict any third label at all; \textsc{Readable} predicts a `neutral' class for $>16\%$ of the text by non-binary authors and is the most `fair' amongst the three platforms. As these softwares are easily available through their web platforms, bias propagation, and reinforcement are very easy. Our study has shown that these platforms are highly biased against non-binary individuals, and any use in downstream tasks without proper acknowledgement of these shortcomings may lead to discriminatory practices against these minority groups. Even if the platforms acknowledge their low accuracies for datasets like social media comments, they fail to identify the bigger social issue of a missing class - viz. non-binary authors. This answers \textbf{RQ2}. Here we see that the performance on the two datasets is similar (from the bar plots in Figure~\ref{fig:existing_plat_accuracy}), answering \textbf{RQ3}.

Finally, we fine-tune a simple BERT-based multilabel classifier to address the shortcomings identified in our audit study. We assessed multiple fine-tuning settings using one or both datasets and tested on the same and the other dataset. Our results showed that the model when fine-tuned on Reddit and tested on the same gave an overall accuracy of 84\%. The non-binary authors' comments were accurately identified 93\% of the time (Figure~\ref{fig:bert_fine_tuned_same_tested_same}). The performance on the Tumblr dataset on the other hand is not as high -- 67\%. We also note that the models do not generalize well unless they have been fine-tuned with data from both platforms in some way. If a model fine-tuned on dataset A is tested for dataset B, the accuracy is lower than 40\% (Figure~\ref{fig:bert_fine_tuned_same_tested_diff}). Fine-tuning using a mixture of the two datasets leads to improved performance (Figure~\ref{fig:bert_fine_tuned_mix_tested_mix}). Our final approach is geared toward understanding how such models can perform under realistic scenarios where gender-annotated data from multiple social media platforms is hard to come by. We fine-tune using the data from one of the datasets and then use few-shot learning (between 0 and 5000 examples for each gender group from the other dataset). The model fine-tuned on Tumblr data and with 5000 Reddit examples (per gender group) performs the best with 90\% accuracy for the non-binary class and overall accuracy of 77\% on the test set from Reddit (Figure~\ref{fig:bert_kshot}). The BERT model is better than the existing platforms for the binary gender prediction as seen in Figure~\ref{fig:bert_fine_tuning} (\textbf{RQ1}) and is highly accurate for predicting the non-binary gender both in the standard fine-tuning setting and the few-shot learning scenario, as seen in Figs.~\ref{fig:bert_fine_tuned_same_tested_same} and \ref{fig:bert_kshot} (\textbf{RQ2}). Finally, the model generalizes well when it observes data from both datasets in some combination, but the individual performance on Tumblr dataset is not noteworthy, whereas the platform learns better from the Reddit dataset (\textbf{RQ3}). We attribute this to the language style prevalent in each of these platforms.

On auditing a small sample of comments on \textsc{ChatGPT}, we observe that performance on Reddit is comparable to our best finetuned BERT model whereas the accuracy on Tumblr is $\approx 13\%$ less (Table~\ref{tab:chatgpt}). This indicates that even a highly sophisticated language model does not identify the gender markers on the Tumblr dataset well and further analysis is needed for more generalizable conclusions (\textbf{RQ4}).

Finally, as part of our future work, we plan to increase the scope of this audit to include more open-source and commercial gender analyzers and test their prediction accuracies on more diverse datasets from other sources like Twitter, Facebook, Internet message boards, and other niche platforms\footnote{https://nonbinary.wiki/wiki/Websites\_and\_social\_networks}. Our experiments using the BERT base model has shown merit in using transformer models for this application and we plan to extend this to other neural architectures which may exhibit equivalent or better performance but with a shorter inference time. This would go a long way in reducing the societal bias against non-binary individuals and also reinstating their voice and position in the online sphere of influence.

\bibliographystyle{ieeetr}
\bibliography{main.bib}

\end{document}